\documentclass[prl,twocolumn,a4paper,twoside]{revtex4}
\usepackage{amsmath,amssymb,amsfonts}
\usepackage{graphicx}
\usepackage[english]{babel}

\begin{document}

\title{Kert\'esz Line in the Three-Dimensional Compact U($1$) Lattice
  Higgs Model}

\date{\today}
\author{Sandro Wenzel}
\author{Elmar Bittner}
\author{Wolfhard Janke}
\author{Adriaan M.J. Schakel}
\author{Arwed Schiller}
\affiliation{Institut f\"ur Theoretische Physik, Universit\"at Leipzig,
  Augustusplatz 10/11, D-04109 Leipzig, Germany}

\begin{abstract}
  The three-dimensional lattice Higgs model with compact U($1$) gauge
  symmetry and unit charge is investigated by means of Monte Carlo
  simulations.  The full model with fluctuating Higgs amplitude is
  simulated, and both energy as well as topological observables are
  measured.  The data show a Higgs and a confined phase separated by a
  well-defined phase boundary, which is argued to be caused by
  proliferating vortices.  For fixed gauge coupling, the phase boundary
  consists of a line of first-order phase transitions at small Higgs
  self-coupling, ending at a critical point.  The phase boundary then
  continues as a Kert\'esz line across which thermodynamic quantities
  are nonsingular.  Symmetry arguments are given to support these
  findings.
\end{abstract}

\maketitle

Being one of the few well-understood theories exhibiting charge
confinement, the three-dimensional (3D) pure compact U($1$) gauge theory
plays a central role in the study of deconfinement transitions
\cite{Polyakov}.  An intriguing extension obtains by coupling a scalar
matter field to this confining gauge theory \cite{FradkinShenker}.  The
resulting Higgs model and its extensions have recently attracted
considerable attention also in the condensed matter community as
effective descriptions of quantum critical phenomena
\cite{Senthil:2004}.  In a seminal paper, Fradkin and Shenker
\cite{FradkinShenker} studied the phase diagram of the model in the
London limit, where the Higgs field has a fixed amplitude.  They
concluded that in the case of a Higgs field carrying one unit charge
$q=1$, it is always possible to move from the Higgs region into the
confined region without encountering singularities in local
gauge-invariant observables.  As for the liquid-vapor transition, this
is commonly interpreted as implying that the two ground states do not
constitute distinct phases.  This is supported further by symmetry
considerations \cite{KovnerRosenstein}. In the absence of matter fields,
the 3D pure noncompact U($1$) gauge theory is characterized by a
\textit{global} magnetic U(1) symmetry.  When magnetic monopoles are
introduced, which are pointlike instanton solutions of the
\textit{compact} U($1$) gauge theory, this global symmetry becomes
anomalous and only the discrete subgroup Z of integer numbers survives. 
When the compact gauge theory is subsequently coupled to a Higgs field
carrying charge $q$, the magnetic symmetry is further reduced to the
finite cyclic subgroup Z$_q$ of $q$ elements.  For $q=2$, this recovers
the known result that the model undergoes a continuous phase transition
belonging to the 3D Ising universality class
\cite{FradkinShenker,Bhanot:1981ug}.  For $q=1$, this argument excludes
a continuous phase transition because the group $Z_1$, consisting of
only the unit element, cannot be spontaneously broken. 

In this Letter, we argue that Monte Carlo data show a more refined
picture with two distinct phases separated by a well-defined phase
boundary.  We consider the $q=1$ model with fluctuating Higgs amplitude,
as was done first in Refs.~\cite{Munehisa:1985rb,Obodi:1985uu} on
smaller lattices, and more recently in
Refs.~\cite{Kajantie:1997vc,Chernodub:2004yv} on larger ones.  The
ensuing picture, which turns out to be closely related to the dual
superconductor mechanism of confinement \cite{'tHooft:1977hy},
essentially vindicates the scenario put forward by Einhorn and Savit
\cite{einhorn2}, in which the transition from the Higgs to the confined
phase is triggered by proliferating vortices.  As discussed below, the
nature of this mechanism is consistent with the Fradkin-Shenker result
\cite{FradkinShenker} and the symmetry argument \cite{KovnerRosenstein}. 
Our results are at odds with the vortex string-breaking scenario put
forward by Nagaosa and Lee \cite{Nagaosa:1999vu}, who recently argued
that only the confined phase is present in the compact theory. 
Furthermore, we found no support for a deconfinement phase transition of
the Berezinsky-Kosterlitz-Thouless (BKT) type recently proposed in
Ref.~\cite{kleinert1}, as for sufficiently large lattices we observe no
scaling behavior at all across the phase boundary where the BKT
transition is supposed to arise. 

The compact U($1$) Higgs model is specified by the Euclidean
lattice action $S = S_g + S_\phi$, with the gauge part
\begin{equation}
\label{Sg} 
  S_g = \beta\sum_{x,\mu<\nu} \left[ 1-\cos \theta_{\mu \nu} (x) \right]. 
\end{equation} 
Here, $\beta$ is the inverse gauge coupling, the sum extends over all
lattice sites $x$ and lattice directions $\mu$, and $\theta_{\mu
  \nu}(x)$ denotes the plaquette variable $\theta_{\mu \nu} (x) =
\Delta_\mu \theta_{\nu} (x) - \Delta_\nu\theta_{\mu} (x)$, with the
lattice derivative $\Delta_\nu \theta_{\mu} (x) \equiv \theta_{\mu}
(x+\nu) - \theta_{\mu} (x)$ and the compact link variable $\theta_{\mu}
(x) \in[-\pi,\pi)$.  The matter part of the action $S$ is given by
\begin{eqnarray} 
\label{hopping}
S_\phi \!\! &=& \!\! -\kappa\sum_{x,\mu}\rho(x) \rho(x+\mu) \cos 
\left[\Delta_\mu\varphi(x) - q \theta_{\mu}(x)\right] 
\nonumber \\ &&  + \sum_x \left\{\rho^2(x) + \lambda 
  \left[\rho^2(x)-1\right]^2 \right\},
\end{eqnarray} 
where polar coordinates are chosen to represent the complex Higgs field
$\phi(x) = \rho(x) \mathrm{e}^{\mathrm{i} \varphi(x)}$, with $\varphi(x)
\in[-\pi,\pi)$, $\kappa$ is the hopping parameter, and $\lambda$ the
Higgs self-coupling.  The pure $|\phi|^4$ theory with fluctuating
amplitude, obtained by taking the limit $\beta \to \infty$, was recently
investigated by means of Monte Carlo simulations in
Ref.~\cite{bittner}.  We consider the system in three spacetime
dimensions, taking one of the dimensions to represent (Euclidean) time. 

The precise nature of the phase diagram is investigated numerically by
studying several (gauge-invariant) observables chosen such that the
gauge and matter parts are probed separately.  Following
Ref.~\cite{schiller1}, where the London limit of the model was
considered, the gauge part is studied by measuring the monopole density
$M$, as defined in Ref.~\cite{degrand}, and the Polyakov loop.  Both
observables distinguish a confined from a deconfined phase.  The matter
part of the model is studied by measuring the Higgs amplitude squared
$\rho^2\equiv(1/L^3)\sum_x \rho^2(x)$, where $L$ is the linear size of
the cubic lattice.  This bulk operator distinguishes the Higgs phase
from a disordered one. In addition, the plaquette action (\ref{Sg})
(divided by $3L^3$) and the so-called coslink observable
\begin{equation}  
\label{coslink}
C =  - \frac{1}{3 L^3} \sum_{x,\mu} \cos  \left[\Delta_\mu\varphi(x) - 
q \theta_{\mu}(x) \right]
\end{equation} 
are monitored.  Both Metropolis and heat-bath methods were used to
generate Monte Carlo updates.  Since these local updates become
inefficient in regions of first-order phase transitions, the
multicanonical method \cite{berg1} and reweighting techniques
\cite{ferrenberg} were implemented to access these
regions of phase space.  The simulations were carried out at fixed
$\beta$ on cubic lattices varying in size from $6^3$ to $32^3$, in
extreme cases to $42^3$.  Thermalization of the production runs
typically took $4 \times 10^4$ sweeps of the lattice, while about $10^6$
sweeps were used to collect data, with measurements taken after each
sweep of the lattice.  The maxima of the coslink susceptibility
$\chi_C=L^3 \left( \langle C^2\rangle - \langle C \rangle^2 \right)$,
and histograms, rescaled to equal height have been used to determine the
location of the phase boundary.  We have chosen the coslink
susceptibility to trace out the phase diagram because its peaks are more
pronounced than for the other observables.  We have checked that within
the achieved accuracy, the monopole susceptibility peaks at the same
location as $\chi_C$ does.  Statistical errors were estimated by means
of jackknife binning.  For a detailed description of the algorithms and
their implementation, see Ref.~\cite{wenzel}. 

\begin{figure}
\centering
\includegraphics[width=0.45\textwidth]{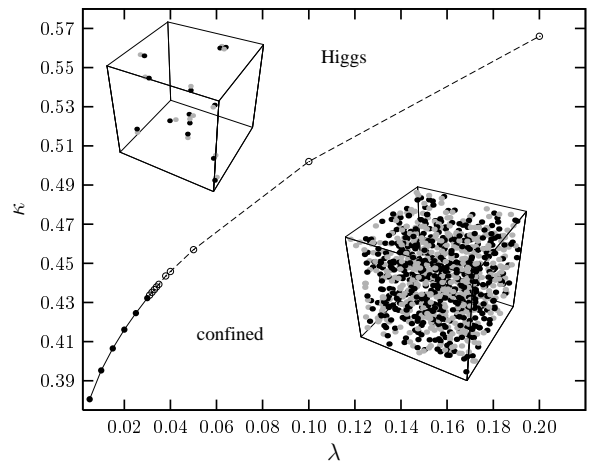}
\caption{\label{fig:phasediagram} Phase diagram of the model at
  $\beta=1.1$ in the infinite-volume limit.  Solid dots mark the
  location of a line of first-order phase transitions.  This line ends
  at a critical point around $0.031 < \lambda_\mathrm{c} < 0.032$.  Open
  dots for $\lambda > \lambda_\mathrm{c}$ mark the location of the
  Kert\'esz line (see text).  Statistical error bars are smaller than
  the symbol size in the figure. The insets show snapshots of typical
  monopole configurations in both phases, with black dots denoting
  monopoles and grey dots denoting antimonopoles.} 
\end{figure}
\begin{figure}
\includegraphics[width=.47\textwidth]{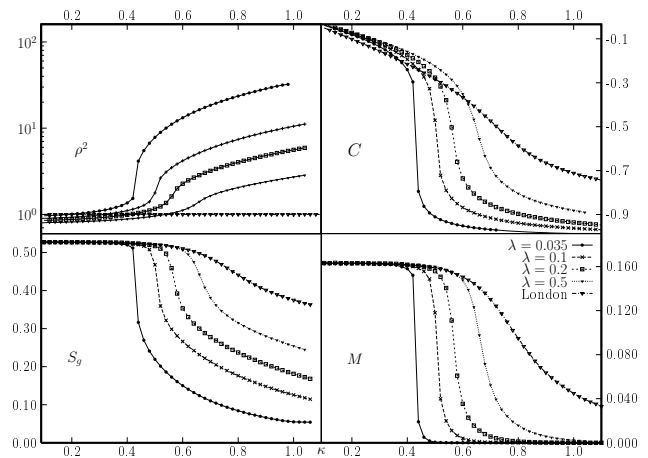}
\caption{\label{fig:observables} Averages of the Higgs amplitude squared
  $\rho^2$, coslink observable $C$, plaquette action $S_g$, and monopole
  density $M$ as a function of $\kappa$ for various values of the
  self-coupling $\lambda$, including the London limit $\lambda \to
  \infty$, at $\beta=1.1$ on a relatively small lattice ($L=12$).} 
\end{figure}

The phase diagram \cite{Munehisa:1985rb,Obodi:1985uu,Chernodub:2004yv},
summarized in Fig.~\ref{fig:phasediagram} for fixed gauge coupling
$\beta=1.1$, consists of two separate phases: a confined and a Higgs
phase.  In the lower right part of the phase diagram, the average Higgs
amplitude squared takes on a minimum value (see
Fig.~\ref{fig:observables}).  The monopole density is finite here and
practically independent of $\kappa$ and $\lambda$.  Snapshots of
monopole configurations (see bottom inset in
Fig.~\ref{fig:phasediagram}) show that the monopoles are in the plasma
phase.  As $\beta$ increases, the monopole density decreases.  The
monopoles become completely suppressed in the weak gauge coupling limit
$\beta \to \infty$, where the model reduces to the pure $|\phi|^4$
theory.  The average plaquette action takes on a value also practically
independent of $\kappa$ and $\lambda$.  These observables signal that
electric charges are confined.  This \textit{confined phase} persists in
the limit $\kappa \to 0$, where the model reduces to the pure compact
U($1$) gauge theory first studied by Polyakov \cite{Polyakov}. Notice
from Fig.~\ref{fig:observables} the non-monotonic behavior of $\langle
\rho^2 \rangle$ and $\langle C \rangle$ as a function of the Higgs
self-coupling $\lambda$.

In the upper left part of the phase diagram, the average $\langle \rho^2
\rangle$ increases more or less linearly with increasing $\kappa$ (note
the logarithmic scale used for $\rho^2$ in Fig.~\ref{fig:observables}). 
The monopole density is vanishing small here and the few monopoles still
present are tightly bound in monopole-antimonopole pairs
\cite{schiller1} (see top inset in Fig.~\ref{fig:phasediagram}).  Being
rendered ineffective, the monopoles can no longer confine electric
charges.  Taken together, these observables identify this phase as
\textit{Higgs phase}.  The identification of the two phases agrees with
the behavior of the Polyakov loop we observed. 

We next examine how for given inverse gauge coupling $\beta$ the
confined phase goes over into the Higgs phase.  Below a critical point
$\lambda_\mathrm{c}(\beta)$, we observe metastable behavior typical for
first-order phase transitions, in accord with earlier Monte Carlo
results obtained on smaller lattices
\cite{Munehisa:1985rb,Obodi:1985uu}.  Simulations for different values
of $\beta \in [1.1, 2.0]$ show that the first-order phase transition
becomes more pronounced for strong gauge coupling, i.e., small $\beta$. 
At fixed $\beta$, the transition becomes stronger with decreasing
$\lambda$, where fluctuations in the Higgs amplitude become more
volatile \cite{Munehisa:1985rb}.  For each value of $\lambda$
considered, the model was simulated on lattices of different sizes to
study finite-size effects and to obtain precise estimates of the
location of the first-order phase transitions, using the multicanonical
approach and reweighting techniques.  The first-order line ends at a
critical point, which for $\beta =1.1$ and infinite volume we estimated
to be located in the interval $0.030 < \lambda_\mathrm{c} < 0.032$.  We
have not attempted to establish the nature of the critical point as
critical slowing down requires very long runs of our code based on
locale updates. 

\begin{figure}
\includegraphics[width=.4\textwidth]{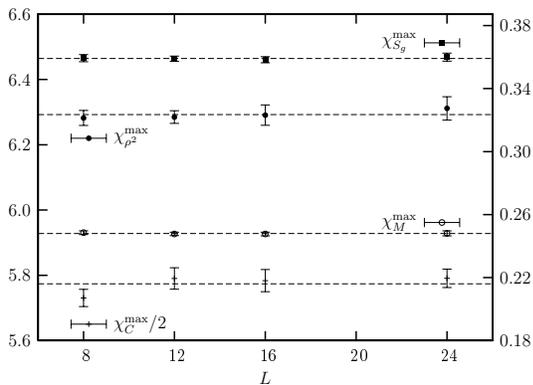}
\caption{\label{fig:sus} Maxima in the susceptibilities of the plaquette
  action $S_g$, Higgs amplitude squared $\rho^2$, monopole density $M$,
  and coslink variable $C$ as a function of the lattice size $L$ at
  $\beta=1.1$ and $\lambda=0.2$.  To fit all the data in one figure,
  $\frac{1}{2}\chi_C^\mathrm{max}$ rather than $\chi_C^\mathrm{max}$ is
  plotted.} 
\end{figure}

Above $\lambda_\mathrm{c}$, we observe the same remarkable behavior as
previously found in the London limit $\lambda \to \infty$, where
fluctuations in the Higgs amplitude are completely frozen
\cite{Bhanot:1981ug,schiller1}.  Namely, for sufficiently large
lattices, the maxima of the susceptibilities do not show any finite-size
scaling (see Fig.~\ref{fig:sus}) and the susceptibility data for the
observables in Fig.~\ref{fig:observables} obtained on different lattice
sizes collapse onto single curves without rescaling, indicating that the
infinite-volume limit is reached.  Since a first-order phase transition
can be excluded in this region, the absence of finite-size scaling
suggests the absence of thermodynamic singularities in the
infinite-volume limit.  To also exclude a crossover in the usual sense,
we have checked that the maxima of $\chi_C$ do not depend on the
direction in which the phase boundary is crossed, either by varying
$\lambda$, or $\kappa$ \cite{Forcrand}.  This analysis is performed using
multihistogram reweighting techniques to achieve a high accuracy in
determining the peak locations.  To sum up this part, despite the
presence of a well-defined and precisely located phase boundary, no
ordinary phase transition or crossover in the usual sense seem to come
into question above $\lambda_\mathrm{c}$. 

In the dual superconductor scenario of confinement
\cite{'tHooft:1977hy}, monopoles are pictured in the Higgs phase as
being tightly bound together in monopole-antimonopole pairs, just as we
observed (see top inset in Fig.~\ref{fig:phasediagram}).  The magnetic
flux emanating from a monopole is squeezed into a short flux tube, or
vortex, carrying one unit $2\pi/q$ ($q=1$) of magnetic flux, which
terminates at an anti-monopole.  The vortices, which in this phase can
also exist as small fluctuating loops, have a finite line tension.  Upon
approaching the phase boundary, the vortex line tension vanishes.  At
this point, the vortices proliferate, gaining configurational entropy
without energy cost, and an infinite vortex network appears which
disorders the Higgs ground state.  At the same time, the monopoles are
no longer bound in tight monopole-antimonopole pairs but form, as seen
in the bottom inset in Fig.~\ref{fig:phasediagram}, a plasma which exhibits
charge confinement.  This scenario explains the existence of a
well-defined phase boundary separating the Higgs and confined phases
\cite{einhorn2}.  The proliferation of vortices persists in the weak
gauge coupling limit $\beta \to \infty$, corresponding to the pure
$|\phi|^4$ theory with global U($1$) symmetry.  Only in this limit, the
proliferating vortices cause a continuous phase transition, belonging to
the XY universality class in which the spontaneously broken global
U($1$) symmetry is restored.  Outside this limit, in the absence of a
relevant global symmetry, the proliferating vortices do not lead to
singularities in thermodynamic quantities and are not connected to a
symmetry breaking transition. 

The situation is reminiscent of the Ising model in an external magnetic
field.  It was shown by Fortuin and Kasteleyn (FK)
\cite{FrotuinKasteleyn} that for zero field this spin model and its
thermal critical behavior can be equivalently formulated as a correlated
percolation problem by putting bonds between nearest neighbor spins in
the same spin state.  The bonds are set with a temperature-dependent
probability $p_\mathrm{FK}(T)$.  The clusters thus constructed percolate
precisely at the Curie point and have the Ising critical exponents
encoded in their fractal structure.  By applying an external field, one
explicitly breaks the global Z$_2$ symmetry of the Ising model, and the
partition function becomes analytic in temperature, excluding a thermal
phase transition.  Yet, for a given applied field $H$, the FK clusters
still percolate at a precisely defined temperature $T_\mathrm{p}(H)$. 
The resulting percolation line in the phase diagram is known as the
\textit{Kert\'esz line} \cite{Kertesz}.  Although percolation
observables remain singular along the line, no thermodynamic
singularities are encountered when crossing it
\cite{Kertesz,Fortunato:2001tx}.  In the limit $H\to \infty$, all the
spins are aligned along the field, so that the FK construction reduces
to random bond percolation.  The Kert\'esz line therefore ends at the
temperature determined by $p_\mathrm{FK}(T_\mathrm{p})=p_\mathrm{c}$,
with $p_\mathrm{c}$ denoting the random bond percolation threshold. 
Along the entire Kert\'esz line, the percolation observables have the
usual percolation exponents. 

The vortex proliferation line in the compact U($1$) Higgs model is the
analog of the Kert\'esz line in the Ising model in an external field.
Such an interpretation of a deconfinement transition as a Kert\'esz line
was first proposed in the context of the SU($2$) Higgs model
\cite{Fortunato:2000ge,Satz:2001zf,Langfeld:2002ic,Bertle:2003pj}.  The
similarity with the Ising model can be made more precise by considering
the London limit $\lambda \to \infty$ of the compact U($1$) Higgs model.
In this limit, the vortex proliferation line starts at the XY critical
point $\kappa=\kappa_\mathrm{XY}, \beta=\infty$.  To identify the
endpoint, we note that for $\kappa \to \infty$, the model reduces to a
Z$_q$ gauge theory.  In 3D, such a discrete gauge theory is dual to the
$q$-state Potts model with \textit{global} Z$_q$ symmetry, which
undergoes a phase transition at some critical value $\beta_q$ of the
inverse gauge coupling \cite{Svetitsky:1985ye}.  Since the limit $q \to
1$ of the Potts model describes random bond percolation, we conclude
that the vortex proliferation line ends at a random bond percolation
critical point at $\beta= \beta_1, \kappa=\infty$.  The vortex network
present in the vicinity of this critical point is expected to be similar
to the one studied in the context of the Kibble mechanism for cosmic
string formation \cite{Vachaspati:1991tr}, which was shown to belong to
the random percolation universality class \cite{Hindmarsh:1994ae}.  By
continuity, we expect that, although not connected to thermodynamic
singularities, the percolation observables have random percolation
exponents along the entire vortex proliferation line.  In the limit
$\beta \to \infty$, these exponents are expected to cross over to the
ones appropriate for the XY universality class.  In a future study, we
plan to investigate the vortex network using percolation observables to
numerically verify these conjectures directly.  Such a study is more
complicated than the indirect study of the vortex network presented
here, via monopoles, which involves only the gauge sector of the theory
and has the advantage that the relevant observables (monopole density
and corresponding susceptibility) can, in contrast to vortex percolation
observables \cite{Kajantie:2000cw,bittner3}, be defined unambiguously.

In conclusion, our Monte Carlo data on the 3D compact U($1$) lattice
Higgs model show that the Higgs and confined phases are separated by a
well-defined phase boundary due to proliferating vortices.  For fixed gauge
coupling, the phase boundary is a line of first-order phase transitions
at small Higgs self-coupling, which ends at a critical point.  The phase
boundary then continues as a Kert\'esz line across which thermodynamic
quantities and other local gauge-invariant observables are nonsingular. 
In the London limit, the Kert\'esz line defined by the proliferating
vortices connects the XY and random bond percolation critical points,
which both form limiting cases of the compact U($1$) Higgs model. 

\vspace{-1mm}
  This work is partially supported by the DFG Grant No. JA 483/17-3. 
  S.W.\ acknowledges financial support from the Studienstiftung des
  deutschen Volkes.  
\vspace{-2mm}
%

\begin{thebibliography}{33}
\expandafter\ifx\csname natexlab\endcsname\relax\def\natexlab#1{#1}\fi
\expandafter\ifx\csname bibnamefont\endcsname\relax
  \def\bibnamefont#1{#1}\fi
\expandafter\ifx\csname bibfnamefont\endcsname\relax
  \def\bibfnamefont#1{#1}\fi
\expandafter\ifx\csname citenamefont\endcsname\relax
  \def\citenamefont#1{#1}\fi
\expandafter\ifx\csname url\endcsname\relax
  \def\url#1{\texttt{#1}}\fi
\expandafter\ifx\csname urlprefix\endcsname\relax\def\urlprefix{URL }\fi
\providecommand{\bibinfo}[2]{#2}
\providecommand{\eprint}[2][]{\url{#2}}

\bibitem[{\citenamefont{Polyakov}(1977)}]{Polyakov}
\bibinfo{author}{\bibfnamefont{A.~M.} \bibnamefont{Polyakov}},
  \bibinfo{journal}{Nucl. Phys.} \textbf{\bibinfo{volume}{B120}},
  \bibinfo{pages}{429} (\bibinfo{year}{1977}). 

\bibitem[{\citenamefont{{E. Fradkin} and {S. H. 
  Shenker}}(1979)}]{FradkinShenker}
\bibinfo{author}{\bibnamefont{{E. Fradkin}}} \bibnamefont{and}
  \bibinfo{author}{\bibnamefont{{S. H. Shenker}}}, \bibinfo{journal}{Phys. 
  Rev.} \textbf{\bibinfo{volume}{D19}}, \bibinfo{pages}{3682}
  (\bibinfo{year}{1979}). 

\bibitem[{\citenamefont{Senthil et~al.}(2004)\citenamefont{Senthil, Balents,
  Sachdev, Vishwanath, and Fisher}}]{Senthil:2004}
\bibinfo{author}{\bibfnamefont{T.}~\bibnamefont{Senthil}},
  \bibinfo{author}{\bibfnamefont{L.}~\bibnamefont{Balents}},
  \bibinfo{author}{\bibfnamefont{S.}~\bibnamefont{Sachdev}},
  \bibinfo{author}{\bibfnamefont{A.}~\bibnamefont{Vishwanath}},
  \bibnamefont{and} \bibinfo{author}{\bibfnamefont{M.~P.~A.} 
  \bibnamefont{Fisher}}, \bibinfo{journal}{Science}
  \textbf{\bibinfo{volume}{303}}, \bibinfo{pages}{1490} (\bibinfo{year}{2004}). 

\bibitem[{\citenamefont{{A. Kovner} and {B. 
  Rosenstein}}(1992)}]{KovnerRosenstein}
\bibinfo{author}{\bibnamefont{{A. Kovner}}} \bibnamefont{and}
  \bibinfo{author}{\bibnamefont{{B. Rosenstein}}}, \bibinfo{journal}{Int. J. 
  Mod. Phys.} \textbf{\bibinfo{volume}{A7}}, \bibinfo{pages}{7419}
  (\bibinfo{year}{1992}). 

\bibitem[{\citenamefont{Bhanot and Freedman}(1981)}]{Bhanot:1981ug}
\bibinfo{author}{\bibfnamefont{G.}~\bibnamefont{Bhanot}} \bibnamefont{and}
  \bibinfo{author}{\bibfnamefont{B.~A.} \bibnamefont{Freedman}},
  \bibinfo{journal}{Nucl. Phys.} \textbf{\bibinfo{volume}{B190}},
  \bibinfo{pages}{357} (\bibinfo{year}{1981}). 

\bibitem[{\citenamefont{Munehisa}(1985)}]{Munehisa:1985rb}
\bibinfo{author}{\bibfnamefont{Y.}~\bibnamefont{Munehisa}},
  \bibinfo{journal}{Phys. Lett.} \textbf{\bibinfo{volume}{B155}},
  \bibinfo{pages}{159} (\bibinfo{year}{1985}). 

\bibitem[{\citenamefont{Obodi}(1986)}]{Obodi:1985uu}
\bibinfo{author}{\bibfnamefont{G.~N.} \bibnamefont{Obodi}},
  \bibinfo{journal}{Phys. Lett.} \textbf{\bibinfo{volume}{B174}},
  \bibinfo{pages}{208} (\bibinfo{year}{1986}). 

\bibitem[{\citenamefont{Kajantie et~al.}(1998)\citenamefont{Kajantie,
  Karjalainen, Laine, and Peisa}}]{Kajantie:1997vc}
\bibinfo{author}{\bibfnamefont{K.}~\bibnamefont{Kajantie}},
  \bibinfo{author}{\bibfnamefont{M.}~\bibnamefont{Karjalainen}},
  \bibinfo{author}{\bibfnamefont{M.}~\bibnamefont{Laine}}, \bibnamefont{and}
  \bibinfo{author}{\bibfnamefont{J.}~\bibnamefont{Peisa}},
  \bibinfo{journal}{Phys. Rev.} \textbf{\bibinfo{volume}{B57}},
  \bibinfo{pages}{3011} (\bibinfo{year}{1998}). 

\bibitem[{\citenamefont{Chernodub et~al.}(2004)\citenamefont{Chernodub,
  Feldmann, Ilgenfritz, and Schiller}}]{Chernodub:2004yv}
\bibinfo{author}{\bibfnamefont{M.~N.} \bibnamefont{Chernodub}},
  \bibinfo{author}{\bibfnamefont{R.}~\bibnamefont{Feldmann}},
  \bibinfo{author}{\bibfnamefont{E.-M.} \bibnamefont{Ilgenfritz}},
  \bibnamefont{and} \bibinfo{author}{\bibfnamefont{A.}~\bibnamefont{Schiller}},
  \bibinfo{journal}{Phys. Rev.} \textbf{\bibinfo{volume}{D70}},
  \bibinfo{pages}{074501} (\bibinfo{year}{2004}). 

\bibitem[{\citenamefont{'t~Hooft}(1978)}]{'tHooft:1977hy}
\bibinfo{author}{\bibfnamefont{G.}~\bibnamefont{'t~Hooft}},
  \bibinfo{journal}{Nucl. Phys.} \textbf{\bibinfo{volume}{B138}},
  \bibinfo{pages}{1} (\bibinfo{year}{1978}). 

\bibitem[{\citenamefont{{M. B. Einhorn} and {R. Savit}}(1979)}]{einhorn2}
\bibinfo{author}{\bibnamefont{{M. B. Einhorn}}} \bibnamefont{and}
  \bibinfo{author}{\bibnamefont{{R. Savit}}}, \bibinfo{journal}{Phys. Rev.} 
  \textbf{\bibinfo{volume}{D19}}, \bibinfo{pages}{1198} (\bibinfo{year}{1979}). 

\bibitem[{\citenamefont{Nagaosa and Lee}(2000)}]{Nagaosa:1999vu}
\bibinfo{author}{\bibfnamefont{N.}~\bibnamefont{Nagaosa}} \bibnamefont{and}
  \bibinfo{author}{\bibfnamefont{P.~A.} \bibnamefont{Lee}},
  \bibinfo{journal}{Phys. Rev.} \textbf{\bibinfo{volume}{B61}},
  \bibinfo{pages}{9166} (\bibinfo{year}{2000}). 

\bibitem[{\citenamefont{{H. Kleinert} et~al.}(2002)\citenamefont{{H. Kleinert},
  {F. S. Nogueira}, and {A. Sudb\o}}}]{kleinert1}
\bibinfo{author}{\bibnamefont{{H. Kleinert}}},
  \bibinfo{author}{\bibnamefont{{F. S. Nogueira}}}, \bibnamefont{and}
  \bibinfo{author}{\bibnamefont{{A. Sudb\o}}}, \bibinfo{journal}{Phys. Rev. 
  Lett.} \textbf{\bibinfo{volume}{88}}, \bibinfo{pages}{232001}
  (\bibinfo{year}{2002}). 

\bibitem[{\citenamefont{{E. Bittner} and {W. Janke}}(2002)}]{bittner}
\bibinfo{author}{\bibnamefont{{E. Bittner}}} \bibnamefont{and}
  \bibinfo{author}{\bibnamefont{{W. Janke}}}, \bibinfo{journal}{Phys.\ Rev.\
  Lett.} \textbf{\bibinfo{volume}{89}}, \bibinfo{pages}{130201}
  (\bibinfo{year}{2002}); \bibinfo{journal}{Phys.\ Rev.} 
  \textbf{\bibinfo{volume}{B71}}, \bibinfo{pages}{024512}
  (\bibinfo{year}{2005}). 

\bibitem[{\citenamefont{{M. N. Chernodub} et~al.}(2002)\citenamefont{{M. N. 
  Chernodub}, {E.-M. Ilgenfritz}, and {A. Schiller}}}]{schiller1}
\bibinfo{author}{\bibnamefont{{M. N. Chernodub}}},
  \bibinfo{author}{\bibnamefont{{E.-M. Ilgenfritz}}}, \bibnamefont{and}
  \bibinfo{author}{\bibnamefont{{A. Schiller}}}, \bibinfo{journal}{Phys.\
  Lett.} \textbf{\bibinfo{volume}{B547}}, \bibinfo{pages}{269}
  (\bibinfo{year}{2002}). 

\bibitem[{\citenamefont{{T. A. DeGrand} and {D. Toussaint}}(1980)}]{degrand}
\bibinfo{author}{\bibnamefont{{T. A. DeGrand}}} \bibnamefont{and}
  \bibinfo{author}{\bibnamefont{{D. Toussaint}}}, \bibinfo{journal}{Phys.\
  Rev.} \textbf{\bibinfo{volume}{D22}}, \bibinfo{pages}{2478}
  (\bibinfo{year}{1980}). 

\bibitem[{\citenamefont{{B. A. Berg} and {T. Neuhaus}}(1992)}]{berg1}
\bibinfo{author}{\bibnamefont{{B. A. Berg}}} \bibnamefont{and}
  \bibinfo{author}{\bibnamefont{{T. Neuhaus}}}, \bibinfo{journal}{Phys. Rev. 
  Lett.} \textbf{\bibinfo{volume}{68}}, \bibinfo{pages}{9}
  (\bibinfo{year}{1992}). 

\bibitem[{\citenamefont{{A. M. Ferrenberg} and {R. H.
  Swendsen}}(1988)}]{ferrenberg}
\bibinfo{author}{\bibnamefont{{A. M. Ferrenberg}}} \bibnamefont{and}
  \bibinfo{author}{\bibnamefont{{R. H. Swendsen}}}, \bibinfo{journal}{Phys. Rev. 
  Lett.} \textbf{\bibinfo{volume}{61}}, \bibinfo{pages}{2635}
  (\bibinfo{year}{1988}); \bibinfo{journal}{Phys. Rev. 
  Lett.} \textbf{\bibinfo{volume}{63}}, \bibinfo{pages}{1195}
  (\bibinfo{year}{1989}). 

\bibitem[{\citenamefont{Wenzel}(2004)}]{wenzel}
\bibinfo{author}{\bibfnamefont{S.}~\bibnamefont{Wenzel}}, Master's thesis,
  \bibinfo{school}{Universit\"at Leipzig} (\bibinfo{year}{2004}). 

\bibitem{Forcrand} We kindly thank P. de Forcrand for suggesting this
  test. 

\bibitem[{\citenamefont{Fortuin and Kasteleyn}(1972)}]{FrotuinKasteleyn}
\bibinfo{author}{\bibfnamefont{C.~M.} \bibnamefont{Fortuin}} \bibnamefont{and}
  \bibinfo{author}{\bibfnamefont{P.~W.} \bibnamefont{Kasteleyn}},
  \bibinfo{journal}{Physica} \textbf{\bibinfo{volume}{57}},
  \bibinfo{pages}{536} (\bibinfo{year}{1972}). 

\bibitem[{\citenamefont{Kert\'esz}(1989)}]{Kertesz}
\bibinfo{author}{\bibfnamefont{J.}~\bibnamefont{Kert\'esz}},
  \bibinfo{journal}{Physica} \textbf{\bibinfo{volume}{A161}},
  \bibinfo{pages}{58} (\bibinfo{year}{1989}). 

\bibitem[{\citenamefont{Fortunato and
  Satz}(2001{\natexlab{a}})}]{Fortunato:2001tx}
\bibinfo{author}{\bibfnamefont{S.}~\bibnamefont{Fortunato}} \bibnamefont{and}
  \bibinfo{author}{\bibfnamefont{H.}~\bibnamefont{Satz}},
  \bibinfo{journal}{Phys. Lett.} \textbf{\bibinfo{volume}{B509}},
  \bibinfo{pages}{189} (\bibinfo{year}{2001}{\natexlab{a}}). 

\bibitem[{\citenamefont{Fortunato and
  Satz}(2001{\natexlab{b}})}]{Fortunato:2000ge}
\bibinfo{author}{\bibfnamefont{S.}~\bibnamefont{Fortunato}} \bibnamefont{and}
  \bibinfo{author}{\bibfnamefont{H.}~\bibnamefont{Satz}},
  \bibinfo{journal}{Nucl. Phys.} \textbf{\bibinfo{volume}{A681}},
  \bibinfo{pages}{466} (\bibinfo{year}{2001}{\natexlab{b}}). 

\bibitem[{\citenamefont{Langfeld}(2002)}]{Langfeld:2002ic}
\bibinfo{author}{\bibfnamefont{K.}~\bibnamefont{Langfeld}}
  (\bibinfo{year}{2002}), \eprint{hep-lat/0212032}. 

\bibitem[{\citenamefont{Bertle et~al.}(2004)\citenamefont{Bertle, Faber,
  Greensite, and Olejnik}}]{Bertle:2003pj}
\bibinfo{author}{\bibfnamefont{R.}~\bibnamefont{Bertle}},
  \bibinfo{author}{\bibfnamefont{M.}~\bibnamefont{Faber}},
  \bibinfo{author}{\bibfnamefont{J.}~\bibnamefont{Greensite}},
  \bibnamefont{and} \bibinfo{author}{\bibfnamefont{S.}~\bibnamefont{Olejnik}},
  \bibinfo{journal}{Phys. Rev.} \textbf{\bibinfo{volume}{D69}},
  \bibinfo{pages}{014007} (\bibinfo{year}{2004}). 

\bibitem[{\citenamefont{Satz}(2002)}]{Satz:2001zf}
\bibinfo{author}{\bibfnamefont{H.}~\bibnamefont{Satz}},
  \bibinfo{journal}{Comput. Phys. Commun.} \textbf{\bibinfo{volume}{147}},
  \bibinfo{pages}{46} (\bibinfo{year}{2002}). 

\bibitem[{\citenamefont{Svetitsky}(1986)}]{Svetitsky:1985ye}
\bibinfo{author}{\bibfnamefont{B.}~\bibnamefont{Svetitsky}},
  \bibinfo{journal}{Phys. Rept.} \textbf{\bibinfo{volume}{132}},
  \bibinfo{pages}{1} (\bibinfo{year}{1986}). 

\bibitem[{\citenamefont{Vachaspati}(1991)}]{Vachaspati:1991tr}
\bibinfo{author}{\bibfnamefont{T.}~\bibnamefont{Vachaspati}},
  \bibinfo{journal}{Phys. Rev.} \textbf{\bibinfo{volume}{D44}},
  \bibinfo{pages}{3723} (\bibinfo{year}{1991}). 

\bibitem[{\citenamefont{Hindmarsh and Strobl}(1995)}]{Hindmarsh:1994ae}
\bibinfo{author}{\bibfnamefont{M.}~\bibnamefont{Hindmarsh}} \bibnamefont{and}
  \bibinfo{author}{\bibfnamefont{K.}~\bibnamefont{Strobl}},
  \bibinfo{journal}{Nucl. Phys.} \textbf{\bibinfo{volume}{B437}},
  \bibinfo{pages}{471} (\bibinfo{year}{1995}). 

\bibitem[{\citenamefont{Kajantie et~al.}(2000)\citenamefont{Kajantie, Laine,
  Neuhaus, Rajantie, and Rummukainen}}]{Kajantie:2000cw}
\bibinfo{author}{\bibfnamefont{K.}~\bibnamefont{Kajantie}},
  \bibinfo{author}{\bibfnamefont{M.}~\bibnamefont{Laine}},
  \bibinfo{author}{\bibfnamefont{T.}~\bibnamefont{Neuhaus}},
  \bibinfo{author}{\bibfnamefont{A.}~\bibnamefont{Rajantie}}, \bibnamefont{and}
  \bibinfo{author}{\bibfnamefont{K.}~\bibnamefont{Rummukainen}},
  \bibinfo{journal}{Phys. Lett.} \textbf{\bibinfo{volume}{B482}},
  \bibinfo{pages}{114} (\bibinfo{year}{2000}). 

\bibitem[{\citenamefont{Bittner et~al.}(2005)\citenamefont{Bittner, Krinner,
  and Janke}}]{bittner3}
\bibinfo{author}{\bibfnamefont{E.}~\bibnamefont{Bittner}},
  \bibinfo{author}{\bibfnamefont{A.}~\bibnamefont{Krinner}}, \bibnamefont{and}
  \bibinfo{author}{\bibfnamefont{W.}~\bibnamefont{Janke}}
  (\bibinfo{year}{2005}), \bibinfo{note}{unpublished}. 

\end{thebibliography}

\end{document}